\documentclass[a4paper,11pt]{article}
\usepackage[margin=2.5cm]{geometry}
\usepackage[utf8]{inputenc}
\usepackage{graphicx}
\usepackage{epsfig}
\usepackage{amsmath}
\usepackage{amssymb}
\usepackage{cite}
\usepackage{float}
\usepackage{subfigure}
\usepackage[font={footnotesize,it}]{caption}
\usepackage{authblk}
\numberwithin{equation}{section}
\usepackage{hyperref}
\usepackage{epstopdf}
\DeclareGraphicsExtensions{.eps}
\usepackage{url}
\begin{document}

\begin{titlepage}

\title{Thermodynamic Geometry of Black Holes in the Canonical Ensemble}
\author[1]{Pankaj Chaturvedi\thanks{\noindent  E-mail:~ cpankaj1@gmail.com}}
\author[2]{Sayid Mondal  \thanks{\noindent E-mail:~   sayidphy@iitk.ac.in}}
\author[3]{Gautam Sengupta\thanks{\noindent E-mail:~  sengupta@iitk.ac.in}}
\affil[1]{
Yau Mathematical Sciences Center\\

Tsinghua University\\

Beijing 100084\\

China
}

\affil[2,3]{
Department of Physics\\

Indian Institute of Technology Kanpur\\

Kanpur 208016\\

India}

\maketitle

\abstract{
\noindent
We investigate the thermodynamics and critical phenomena for four dimensional RN-AdS and Kerr-AdS black holes 
in the canonical ensemble both for the normal and the extended phase space employing the framework of thermodynamic geometry. The thermodynamic scalar curvatures for these black holes characterize the liquid-gas like first order phase transition analogous to the van der Waals fluids, through the $R$-Crossing Method. It is also shown that the thermodynamic scalar curvatures diverge as a function of the temperature at the critical point.

}
\end{titlepage}
\tableofcontents
\newpage
\section{Introduction}
Over the last several decades the area of black hole thermodynamics has witnessed significant developments
making it an important testing ground for candidate theories of quantum gravity \cite{Wald:1999vt,Page:2004xp, hawking1983thermodynamics}. Despite significant progress through diverse frameworks, the exact microscopic statistical structure of black holes is still far from being fully elucidated in the absence of a consistent theory of quantum gravity. However the semi classical thermodynamic approach has led to significant insights into the phase structure and critical phenomena for disparate black holes in arbitrary dimensions. In this context the investigation of the thermodynamics of asymptotically anti-de Sitter (AdS) black holes have assumed a central role owing to the well known $AdS/CFT$ correspondence \cite{maldacena1999large}. Unlike asymptotically flat black holes the latter are thermodynamically stable and exhibit a rich variety of phase transitions and critical phenomena. In particular  the Hawking-Page phase transition in such asymptotically AdS black holes are related to the confinement/deconfinement transition in the boundary field theory through the AdS-CFT correspondence \cite{Witten:1998zw}. For these reasons the study of the phase transitions and critical phenomena of asymptotically AdS black holes has received significant attention over the last two decades 
( see \cite{davies1978thermodynamics,chamblin1999charged,
chamblin1999holography,caldarelli2000thermodynamics,
niu2012critical,tsai2012phase} for comprehensive references).
  
Unlike conventional thermodynamic systems the phase structure of black holes is ensemble dependent as they are locally gravitating configurations. Their entropy is non extensive and conventional thermodynamic stability arguments are not valid. Despite this important distinction asymptotically AdS black holes closely resemble conventional thermodynamic systems. For example the charged Reissner-Nordstrom-AdS (RN-AdS) black holes in a canonical ensemble (fixed charge) exhibit a first order liquid-gas like phase transition (analogous to the van der Waals fluids) culminating in a second order critical point  \cite{chamblin1999holography,chamblin1999charged}. In the grand canonical ensemble (fixed electric potential) however, these black holes undergo a Hawking-Page phase transition \cite{peca1999thermodynamics} to a thermal AdS space time illustrating the ensemble dependence of their phase structure. 

In the recent past investigations in this area have been focussed on an extended thermodynamic phase (state) space through the identification of the cosmological constant $\Lambda$ as a thermodynamic pressure with a corresponding conjugate thermodynamic volume \cite{kastor2009enthalpy, dolan2010cosmological, dolan2011compressibility, cvetivc2011black}. This naturally leads to a modification of the Smarr formula and a consequent additional term in the first law requiring the identification of the ADM mass of the black hole with the enthalpy \cite{kastor2009enthalpy}. Remarkably this approach renders the phase structure of RN-AdS and Kerr-AdS black holes in the canonical ensemble, to be identical to that of the van der Waals fluids with an exact match for the corresponding critical exponents \cite{kubizvnak2012p,rajagopal2014van,gunasekaran2012extended}.  

In a related development, over the last two decades a consistent geometrical framework for studying phase transitions and critical phenomena for thermodynamic systems has received considerable interest. Starting from the pioneering work of Weinhold \cite {weinhold1975metric} and Ruppeiner \cite {ruppeiner1995riemannian} an intrinsic geometrical framework to study thermodynamics and phase transitions have been systematically developed. This approach involves a Euclidean signature Riemannian geometry of thermodynamic fluctuations in the equilibrium state space of the system. Interestingly the positive definite line interval ( with the dimensions of volume) connecting two equilibrium states in this geometry, may be related to the probability distribution of thermodynamic fluctuations in a Gaussian approximation. Subsequently from standard scaling and hyperscaling arguments it could be shown that the interactions in the underlying microscopic statistical basis are encoded in the thermodynamic scalar curvature arising from this geometry. Consequently the thermodynamic scalar curvature scales like the {\it correlation volume} of the system and diverges at the critical point. The scalar curvature was observed to be inversely proportional to the singular part of the free energy arising from long range correlations which vanishes at the critical point \cite{ruppeiner1995riemannian}. Notice that although this geometrical approach is thermodynamic and involves the macroscopic description of the system it serves as a bridge to the microscopic description through the consideration of the Gaussian fluctuations in the analysis. This renders the geometrical framework described above a convenient method for the description of the thermodynamic of black holes where a clear idea of the microscopic structure is still elusive. Over the last decade or so this geometrical framework has provided significant insights into the phase structure  and critical phenomena for black holes \cite{gibbons1996moduli,ferrara1997black,sarkar2006thermodynamic,sarkar2008thermodynamic,cai1999critical,cai1999thermodynamic, ruppeiner2007black,aaman2003geometry,aaman2006flat,gibbons2005first,shen2007thermodynamic,banerjee2012critical,   
banerjee2012critical1}.

Following the above developments one of the authors (GS) in the collaborations 
\cite {sahay2010thermodynamic,sahay2010thermodynamic1,sahay2010thermodynamic2, ruppeiner2012thermodynamic,sahay2016state} established a unified geometrical framework for the characterization of subcritical, critical and supercritical phenomena for thermodynamic systems including black holes. The geometrical characterization of the subcritical first order phase transitions involved the extension of Widom's microscopic approach to phase transitions based on the correlation length \cite {widom1965equation,widom1974critical,stanley1999scaling}. It was proposed that the phase coexistence at a first order phase transition implied the equality of the correlation lengths for phase coexistence at a first order phase transition. For the geometrical framework described above this proposal translated to the crossing of the branches for the multiple valued thermodynamic scalar curvature as a function of its arguments. This was referred to as the {\it R-Crossing Method} and described the equality of the thermodynamic scalar curvature at a first order phase transition and the results for disparate fluid systems exhibited a remarkable correspondence with experimental data \cite {ruppeiner2012thermodynamic,may2012riemannian,may2013thermodynamic,simeoni2010widom,dey2013information}. This characterization has also been demonstrated for the first order phase transition in dyonic charged AdS black hole in a mixed canonical/grand canonical ensemble with a fixed magnetic charge and a varying electric charge of the black hole by two of the authors (PC and GS) in the collaboration \cite{chaturvedi2014thermodynamic}.

Despite the progress describe above the characterization of the phase structure of black holes in the canonical ensemble
through the framework of thermodynamic geometry has remained an unresolved issue. As outlined in \cite {ruppeiner1995riemannian} the form of the thermodynamic metric is crucially dependent on the choice of the correct thermodynamic potential as a function of the thermodynamic variables. This in turn is determined by the choice of the ensemble being considered. In this article we critically examine this unresolved issue and propose a construction for the
thermodynamic geometry of four dimensional RN-AdS and Kerr-AdS black holes in the canonical ensemble in both the extended and the normal phase space. The appropriate thermodynamic potential in this case is the Helmholtz free energy as a function of the corresponding relevant thermodynamic variables. The thermodynamic metric appropriate to the canonical ensemble may then be obtained as the Hessian of the corresponding thermodynamic potential with respect to its arguments.
Interestingly the thermodynamic scalar curvature arising from our proposed geometrical construction correctly characterizes the phase structure and critical phenomena for these black holes in the canonical ensemble, and matches well with the results from the conventional free energy approach. This naturally serves as an important application of the unified geometrical description described earlier, for phase transitions and critical phenomena, developed by one of the authors (GS), for the case of black holes. In particular it is a significant additional confirmation of the {\it $R$-Crossing Method} for first order phase transitions described in \cite {ruppeiner2012thermodynamic, chaturvedi2014thermodynamic}. 
 
This article is organized as follows. In the next two sections we briefly review  the thermodynamics and phase structure of four dimensional RN-AdS and  Kerr-AdS black holes both in the normal and the extended phase space. In section four we briefly present the essential elements of the framework of thermodynamic geometry. In the subsequent sections five and six we describe the construction of the thermodynamic geometry for  RN-AdS and Kerr-AdS black holes in the canonical ensemble and the characterization of their phase structures in this geometrical framework through the thermodynamic scalar curvature. In the concluding section seven we present a summary of our results and future issues.
\section{Thermodynamics of Four Dimensional  RN-AdS Black Holes}
In this section  we briefly review some of the basic thermodynamic properties of four dimensional RN-AdS black holes
which are solutions of the Einstein-Maxwell equations with a negative cosmological constant. The corresponding four dimensional Einstein-Maxwell-AdS action is given as \cite{kubizvnak2012p}
\begin{equation}\label{RNAdS action}
I_{EM}=-\frac{1}{16\pi}\int _{M}\sqrt{-g}\left(R-F^2+\frac{6}{l^2}\right).
\end{equation}
Here $\Lambda=-\frac{6}{l^2}$ is the cosmological constant which is related to the AdS length scale $l$ and $R$ is the Ricci scalar. The RN-AdS metric is spherically symmetric and static and is described as follows,
\begin{equation}
ds^{2}=-fdt^{2}+f^{-1}dr^{2}+r^{2}d\Omega_{2}^{2}.
\end{equation}
The lapse function $f(r)$ is given as
\begin{equation}
f=1- \frac{2M}{r}+\frac{Q^{2}}{r^2}+\frac{r^2}{l^2},
\end{equation}
and the larger root of the equation $f(r)=0$ determines the black hole horizon at $r=r_{+}$.
The field strength for the $U(1)$ Maxwell gauge field $A$ is given as
\begin{equation}\label{field strength of RN}
F=dA,~~~~~~A=-\frac{Q}{r}dt.
\end{equation}
The parameter $M$ and $Q$ are the ADM mass and the electric charge of the black hole respectively and the electric potential $\Phi$ between the horizon and the asymptotic infinity  may be expressed as
\begin{equation}
\Phi=\frac{Q}{r_{+}}.
\end{equation}
The free energy $F$ for the system is related to  the partition function $Z$ as
\begin{equation}
F=-k_{B}T ~ln ~Z .
\end{equation}
In order to determine the partition function $Z$, one computes the Euclidean action $I$ for the system which is given by the following expression,
\begin{equation}
I= I_{EM}+ I_{s}+I_{c},
\end{equation}
where  $I_{EM}$ is the Einstein-Maxwell action Eq.(\ref{RNAdS action}).
The other  two terms in the action, $I_{s}$ and $I_{c}$ are the surface integral for fixed charge and the invariant counter term that is needed to cure the infra-red divergences\cite{mann1999misner,emparan1999surface}. The total action is given as \cite{chamblin1999holography,caldarelli2000thermodynamics,kubizvnak2012p} 
\begin{equation}
I=\frac{\beta}{4l^2}\left(l^2r_{+}-r^3_{+}+\frac{3l^2Q^2}{r_{+}}\right),
\end{equation}
where $\beta$ is the inverse temperature. The Hawking temperature of the black hole is described as
\begin{equation}\label{beta}
T=\frac{1}{\beta}=\frac{1}{4\pi}f^{'}(r_{+})=\frac{1}{4 \pi r_{+}}\left(1+\frac{3r_{+}^2}{l^2}-\frac{Q^2}{r_{+}^2}\right),
\end{equation}
and the entropy of the black hole is  
\begin{equation}
S=\dfrac{A}{4},~~~~~~ A=4\pi r_{+}^2,
\end{equation}
where $A$ is the area of the event horizon.
\subsection{Thermodynamics in the Normal Phase Space}
For the normal phase space  the cosmological constant $\Lambda$ is held fixed and hence the AdS length scale $l$ may be scaled to unity. The Helmholtz free energy $F$ and the temperature $T$ in the canonical ensemble may be described as \cite {chamblin1999charged,chamblin1999holography},
\begin{equation}\label{free energy RNNe}
F=M-TS=\frac{1}{12} \left(\frac{9Q^2}{r_{+}} + 3 r_{+} - r_{+}^3\right),
\end{equation}
 and
\begin{equation}\label{tem RNne}
T=\frac{1}{4 \pi r_{+}}\left(1+3r_{+}^2-\frac{Q^2}{r_{+}^2}\right).
\end{equation}
\begin{figure}[H]
\centering
\includegraphics[scale=.55]{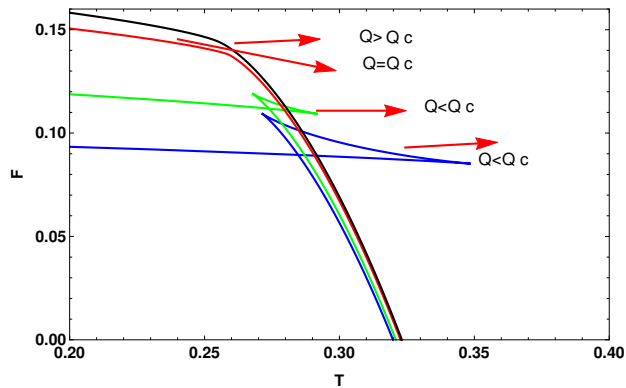}
\caption{\label{RN-AdS1}\textbf{Free energy: RN-AdS black holes in the normal phase space:} $F$ is plotted against $T$ for different value of $Q$. Characteristic swallowtail behaviour is obtained for $Q< Q_{c}=0.17$, corresponding to SBH/LBH first order phase transition. At $Q=Q_{c}$ first order transition culminates in a second order critical point. The corresponding values of the charge $Q$ for the curves are as follows, $Q=0.150$ (blue), $Q=0.130$ (Green), $Q_{c}=0.170$ (red) and $Q=0.175$ (black). }	
\end{figure}
The behaviour of the free energy $F$ with the temperature $T$  is shown in the fig.[\ref{RN-AdS1}] as described in \cite{chamblin1999charged,chamblin1999holography}.  The free energy curves for different charges exhibits a {\it characteristic swallowtail}~  structure  indicating a first order  small-black-hole/large-black-hole \textbf{(SBH/LBH)} phase transition for $Q<Q_{c}$, where $Q_{c}$ is the  value of charge at the critical point. This is analogous to the liquid-gas   phase transition for  van der Waals fluids. It may be observed  from the fig.[\ref{RN-AdS1}] that the blue curve $(Q=0.10)$ and the green curve $(Q=0.13)$ correspond to $Q<Q_c$. These describe three distinct branches that constitute the   {\it swallowtail} structure in the phase diagram. The red curve in the fig.[\ref{RN-AdS1}] corresponds to the value of the critical charge  $Q=Q_{c}=0.17$ which indicates the second order phase transition at the critical point with the corresponding critical temperature as $T_c=0.257$.
\subsection{Thermodynamics in the Extended Phase Space}
In the extended phase space the cosmological constant $\Lambda$ is identified as thermodynamic  pressure and its variation is included in the first law of thermodynamics\cite{kastor2009enthalpy,
dolan2010cosmological,dolan2011compressibility,cvetivc2011black}. Thus we have 
\begin{equation}\label{cosmological}
P=-\frac{1}{8\pi}\Lambda=\frac{3}{8\pi}\frac{1}{l^2}.
\end{equation}
The conjugate variable to the pressure is the thermodynamic volume $V=\left(\frac{\partial M}{\partial P}\right)_{S,Q}$. For the four dimensional RN-AdS black hole the conjugate volume is given as
\begin{equation}
V=\frac{4}{3}\pi r_{+}^3,
\end{equation}
where $r_{+}$ is the radius of the  horizon. In the extended phase space the mass $M$ of the black hole is  identified with the enthalpy rather than the internal energy \cite{kastor2009enthalpy}, and may be expressed as
\begin{equation}
M = \frac{3 Q^2 + 3 r_{+}^2 + 8 P \pi r_{+}^4}{6 r_{+}}.
\end{equation}
The Helmholtz free energy $F$ and the temperature $T$ for the canonical ensemble in the extended phase space are described as \cite{kubizvnak2012p}
\begin{align}
\label{RNEFE}
\begin{split}
F=\frac{1}{4}\left(r_{+}-\frac{8\pi}{3}Pr_{+}^3+\frac{3Q^2}{r_{+}}\right),\\
T=\frac{1}{4\pi}f^{'}(r_{+})=\frac{-Q^2 + r_{+}^2 + 8 P \pi r_{+}^4}{4 \pi r_{+}^3}.
\end{split}
\end{align}

\begin{figure}[H]
\centering
\includegraphics[scale=.6]{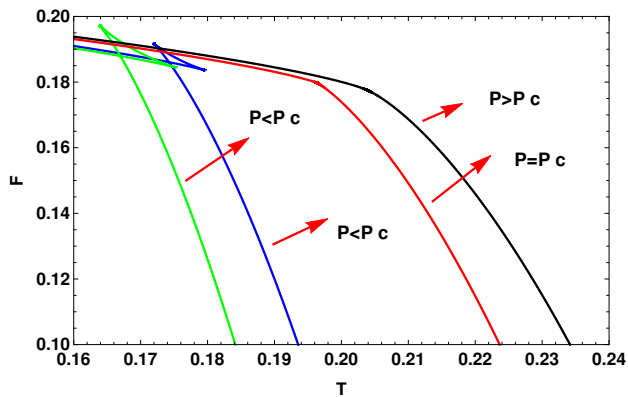}
\caption{\label{rnefe}\textbf{Free energy: RN-AdS black hole in the extended phase space.} $F$ is plotted against $T$ for a fixed $Q=0.22$ and different pressures $P$.  Characteristic swallowtail behaviour is obtained below critical pressure $P<P_{c}=0.068$, indicating  SBH/LBH first order phase transition. The corresponding values of the pressure $P$ are as follows, $P=0.060$ (blue), $P=0.065$ (green), $P_{c}=0.068$ (red) and $P=0.075$ (black).}		
\end{figure}
The behaviour of the free energy  $F$  with the temperature  $T$  for different values of the pressure $P$ and fixed values of the charge $Q$ are shown in fig.[\ref{rnefe}], which was first obtained in\cite{kubizvnak2012p}. As elucidated earlier these $F$ curves exhibit  branched  {\it swallowtail} structure characterizing first order \textbf{SBH/LBH} phase transition analogous to the liquid-gas  phase transition for van der Waals fluids. These are depicted by the blue $(P=0.060,Q=0.22)$ and the green $(P=0.065,Q=0.22)$ curves. It is observed that the (red curve) for the pressure $P=P_{c}=0.068$, where $P_{c}$ is the value of the pressure at the critical point and for a fixed charge $Q=0.220$ culminates in a second order critical point with the corresponding critical temperature as $T_c=0.196$.
\section{Thermodynamics of Four Dimensional  Kerr-AdS Black Holes}
In this section we briefly review the basic thermodynamic properties of four dimensional  Kerr-AdS black holes  and discuss  phase structure of these black holes both in the normal and the extended phase space in the  canonical ensemble.  In the Boyer-Lindquist coordinates the axisymmetric Kerr-AdS metric takes the form \cite{altamirano2013reentrant}
\begin{align}
\label{eqn:eqlabel}
\begin{split}
ds^{2}= -\frac{\Delta}{\rho^{2}} [dt-\frac{a \sin^{2}\theta}{\Xi} d\phi]^{2}+\frac{\rho^{2}}{\Delta}dr^{2}+\frac{\rho^{2}}{\Delta_{\theta}}d\theta^{2}\\ 
+\frac{\Delta_{\theta}\sin^{2}\theta}{\rho^{2}}[a ~dt-\frac{(r^{2}+a^{2})}{\Xi}d\phi ]^{2},
\end{split}
\end{align}
where
\begin{align}
\label{eqn:eqlabel}
\begin{split}
 \rho^{2}&=r^{2}+a^{2}\cos^{2}\theta,~~~~~  \Xi=1-\frac{a^{2}}{l^{2}},\\
  \Delta&=(r^{2}+a^{2})(1+\frac{r^{2}}{l^{2}})-2mr, ~~~ \Delta_{\theta}=1-\frac{a^{2}}{l^{2}}\cos^{2}\theta.
\end{split}
\end{align}
The corresponding thermodynamic quantities are as follows
\begin{align}
\label{temp kerr extended}
\begin{split}
 M&=\frac{m}{\Xi^{2}}=\frac{l^2 (a^2 + r_{+}^2) (l^2 + r_{+}^2)}{2 (a^2 - l^2)^2 r_{+}},~~~ J=\frac{ma}{\Xi^{2}},~~~~~~ \Omega_{H}=\frac{a \Xi}{(r_{+}^{2}+a^{2})},\\
 S&=\frac{\pi (r_{+}^{2}+a^{2})}{  \Xi}, ~~~~~~~~T =\frac{1}{2\pi}\left[r_+\left(\frac{r_+^2}{l^2}+1\right)\left(\frac{1}{a^2 +r_+^2}+\frac{1}{2 r_+}\right)-\frac{1}{r_+}\right].
\end{split}
\end{align}
Free energy $F$ may be obtained as before, which is given as
\begin{equation}\label{free energy Kerr}
F=M-TS=\frac{I}{\beta}+\Omega_{H}J=\frac{r}{4 \Xi^2}\left(3a^2+r_{+}^2-\frac{(r_{+}^2-a^2)^2}{l^2}+\frac{3a^2r_{+}^4+a^4r_{+}^2}{l^4}  \right).
\end{equation}
where $I$ is the Euclidean action\cite{caldarelli2000thermodynamics}.
\subsection{Thermodynamics in the Normal Phase Space}
As described earlier the Helmholtz free energy $F$ and the Hawking temperature $T$ of the Kerr-AdS black holes may be described in a similar fashion as given below \cite {caldarelli2000thermodynamics}
\begin{align}
\label{kerr normal}
\begin{split}
F=&\frac{r_{+}}{4 \Xi^2}\left[3a^2+r_{+}^2-(r_{+}^2-a^2)^2+3a^2r_{+}^4+a^4r_{+}^2  \right],\\
T=&\frac{1}{2\pi}\left[r_+\left(r_+^2+1\right)\left(\frac{1}{a^2 +r_+^2}+\frac{1}{2 r_+}\right)-\frac{1}{r_+}\right].\\
 \end{split}
\end{align}
\begin{figure}[H]
\centering
\includegraphics[scale=.55]{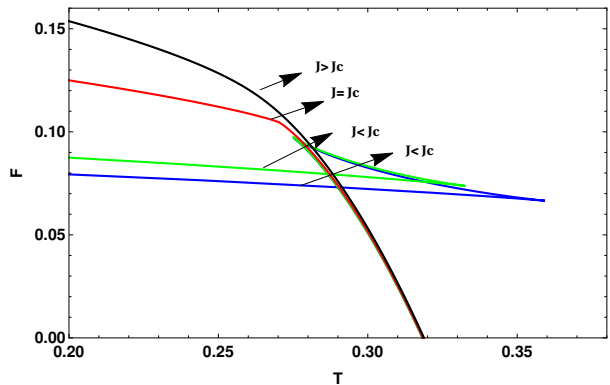}
\caption{\label{plot3} \textbf{Free energy: Kerr-AdS black holes in the normal phase space.} $F$ is plotted against $T$ for different values of angular momentum $J$.  Characteristic swallowtail behaviour is obtained for $J<J_{c}=0.236$, corresponding to  SBH/LBH first order phase transition. The corresponding values of the angular momentum $J$ are as follows, $J=0.01$ (blue), $J=0.005$ (green), $J_{c}=0.0236$ (red) and $J=0.05$ (black).}		
\end{figure}

The behaviour of the Helmholtz free energy  $F$ with the temperature $T$ for different values of the angular momentum $J$ is shown in the fig.[\ref{plot3}], which was first obtained in\cite{caldarelli2000thermodynamics}. As earlier these $F$ curves exhibit branched {\it swallowtail} structure characterizing a first order SBH/LBH phase transition for $J<J_{c}=0.0236$, where $J_{c}$ is the value of the angular momentum at the critical point. These are illustrated  by the blue $(J=0.01)$ and the green $(J=0.005)$ curves  in the fig.[\ref{plot3}]. The red curve in the fig.[\ref {plot3}] corresponds to the value of the critical angular momentum $J=J_{c}=0.0236$, which indicates the second order phase transition at the critical point with the corresponding critical temperature as $T_c=0.270$.
\subsection{Thermodynamics in the Extended Phase Space}
The free energy $F$ and the temperature $T$ in the extended phase for the  canonical ensemble may be reproduced directly from the earlier Eq.(\ref{free energy Kerr}) and [\ref{temp kerr extended}] respectively. These are as follows,
\begin{align}
\label{kerr extended}
\begin{split}
F=&\frac{r}{4 \Xi^2}\left[3a^2+r^2-\frac{8}{3}\pi P(r^2-a^2)^2+(\frac{8}{3}\pi P)^2(3a^2r^4+a^4r^2 )\right],\\
T =&\frac{1}{2\pi}\left[r_+\left(\frac{8\pi P r_+^2}{3}+1\right)\left(\frac{1}{a^2 +r_+^2}+\frac{1}{2 r_+}\right)-\frac{1}{r_+}\right].
 \end{split}
\end{align}
\begin{figure}[H]
\centering
\includegraphics[scale=.55]{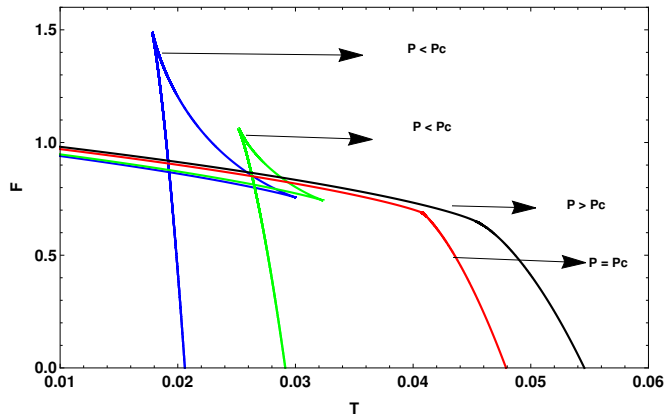}
\caption{\label{plot4}\textbf{Free energy: Kerr-AdS black hole in the extended phase space.} $F$ is plotted for fixed $J=1$. Characteristic swallowtail behaviour is obtained for $P<P_{c}=0.236$, corresponding to SBH/LBH first order phase transition. The corresponding values of the pressure $P$ are as follows, $P=0.0005$ (blue), $P=0.01$ (green), $P_{c}=0.027$ (red) and $P=0.0035$ (black).}		
\end{figure}
The behaviour of the free energy $F$ with the temperature $T$ for different values of the pressure $P$ and a fixed angular momentum  $J$ is depicted in fig.[\ref{plot4}], which was first obtained in\cite{altamirano2013reentrant}. The free energy $F$ curves depicted by  blue $(P=0.0005,J=1)$ and the green $(P=0.01,J=1)$ in the fig.[\ref{plot4}] as earlier illustrates the {\it characteristic swallowtail} structure indicating a first order phase transition for $P<P_c=0.027$, between SBH/LBH phases analogous to the liquid-gas  phase  transition for van der Waals fluids. The red curve in the fig.[\ref {plot4}] corresponds to the value of the critical pressure $P=P_{c}=0.027$ which indicates the second order phase transition at the critical point with the corresponding critical temperature $T_c=0.040$.
\section{Thermodynamic Geometry in the Canonical Ensemble}
In this section we briefly summarize some essential elements of the thermodynamic geometry and its application to study the thermodynamics and the phase structure of black holes. The intrinsic thermodynamic geometrical description of  equilibrium thermodynamic was introduced by Weinhold \cite{weinhold1975metric} in the energy representation. The Weinhold metric may be expressed as the Hessian of the specific internal energy function $u$ ($u=U/V$) with respect to its arguments as,
\begin{equation}
g_{\mu \nu}=\frac{1}{T}\frac{\partial^2 u}{\partial x^{\mu} \partial x^{\nu}},
\end{equation}
where the arguments $x^{\mu}$ are the extensive variables which serve as coordinates in the equilibrium thermodynamic state space. However a clear physical interpretation could not be ascribed to the invariant line element in this geometry. 
Subsequently Ruppenier\cite{ruppeiner1995riemannian} using the entropy representation and the elements of thermodynamic fluctuation theory proposed a metric that was given by Hessian of the specific entropy ($s=S/V$) with respect to the extensive variables as, 
\begin{equation}
g_{\mu \nu}=-\frac{\partial^2 s}{\partial x^{\mu} \partial x^{\nu}}.
\end{equation}
As mentioned earlier in the introduction the probability distribution of thermodynamic fluctuations between two infinitesimally separated
equilibrium states in the thermodynamic state space could then be related to the invariant line element in a Gaussian approximation as 
\begin{equation}
P\left(x+\Delta x, x\right)= A~exp[-\frac{1}{2} g_{\mu \nu}\left(x\right) dx^{\mu}dx^{\nu}],
\end{equation}
where $A$ is a constant. The thermodynamic scalar curvature $R$ for this geometry encoded the interactions in the underlying microscopic statistical system and was zero for non interacting system like the ideal gas.
Through standard scaling and hyperscaling arguments it could be shown that the thermodynamic scalar curvature $R$ scaled as the {\it correlation volume} of the system 
\begin{equation}
R\thicksim \xi^d,
\end{equation}
where $\xi$ and $d$ are  correlation length and   physical dimensionality respectively. From the invariance of the line element under general coordinate transformations it could be shown that any Masieu function obtained through Legendre transformations could serve as a thermodynamic potential for the corresponding thermodynamic metric \cite{ruppeiner1995riemannian}. At a critical point of second order phase transition the correlation length becomes infinite leading to the divergence of the thermodynamic scalar curvature. It could be further shown in \cite{sahay2010thermodynamic,sahay2010thermodynamic1,
 sahay2010thermodynamic2} that the thermodynamic scalar curvature $R$ was a multiple valued function of its arguments at a first order phase transition which could be characterized by the crossing of the branches corresponding to the coexisting phases. This was termed as the {\it $R$-Crossing Method} and provided an alternative geometrical characterization to the usual free energy based condition for first order phase transitions and phase coexistence. Moreover the locus of the maxima of $R$ beyond the critical point provided a direct theoretical scheme to compute the Widom line in the supercritical region leading to a complete unified geometrical characterization of first order phase transitions, critical and supercritical phenomena. The {\it $R$ -Crossing Method} was successfully implemented to describe the first order phase transitions for dyonic charged black holes in mixed canonical/grand canonical ensembles \cite{chaturvedi2014thermodynamic}. In this article we apply this geometrical characterization to study the liquid-gas like phase behaviour of RN-AdS and Kerr-AdS black holes in the  canonical ensemble both in the  normal and  the extended phase space.    
\section{Thermodynamic Geometry of RN-AdS Black Holes }
In this section we provide a construction for  the  thermodynamic geometry of four dimensional RN-AdS black holes in the canonical ensemble for both  the normal and the extended  phase space. To this end we use the Helmholtz free energy $F$ as the thermodynamic potential to obtain the thermodynamic metric which may be expressed as \cite {ruppeiner1995riemannian}
\footnote { Note that black holes have no notion of physical volume hence it is conventional to use the normal thermodynamic potential instead of the specific one in the definition of the thermodynamic metric.}
\begin{equation}\label{metric rn}
g_{\mu \nu}=\frac{1}{T}\frac{\partial^2 F}{\partial x^{\mu} \partial x^{\nu}}.
\end{equation}
Note that in the canonical ensemble the charge $Q$ is held fixed and hence appropriate derivatives with respect to the arguments of the Helmholtz free energy $F$ must be considered \footnote {See also reference \cite {sahay2016state} for an alternate approach in the case of the extended phase space.}. In this case these are the electric potential $\phi$ and the temperature $T$. The scalar curvature $R$ computed from this thermodynamic metric may then be examined as a function of the relevant thermodynamic variable to describe the first order phase transition according to the {\it $R$-Crossing Method} described earlier.  
\subsection{Thermodynamic Geometry in the Normal Phase Space}
In the normal phase space the thermodynamic scalar curvature $R$ is obtained from the metric defined by the Eq.(\ref{metric rn}), where $F$ and $T$ are the free energy and the temperature respectively, which are  given by the Eq. (\ref{free energy RNNe}). The appropriate thermodynamic variables in this case are  the temperature $T$ and the electric potential $\phi$ in the thermodynamic state space $x^{\mu}=\left(T,\phi\right)$. The thermodynamic scalar curvature $R$ may then be expressed as follows 
\begin{align}
\label{eqn:eqlabel}
\begin{split}
R=&\frac{N}{D},\\
N=&2 \left[36 Q^6 - 3 Q^4 r_+^2 (21 + 32 r_+^2) + Q^2 r_+^4 (19 + 64 r_+^2 + 4 r_+^4) -r_+^6 (-4 + r_+^2 + r_+^4 + 8 r_+^6)\right],\\
D=&3 (-Q^2 + r_+^2 + r_+^4) (3 Q^2 r_+ - r_+^3 + r_+^5)^2.
\end{split}
\end{align}
\begin{figure}[H]
 \centering
\begin{minipage}[b]{0.45\linewidth}
\includegraphics[width =2.8in,height=1.8in]{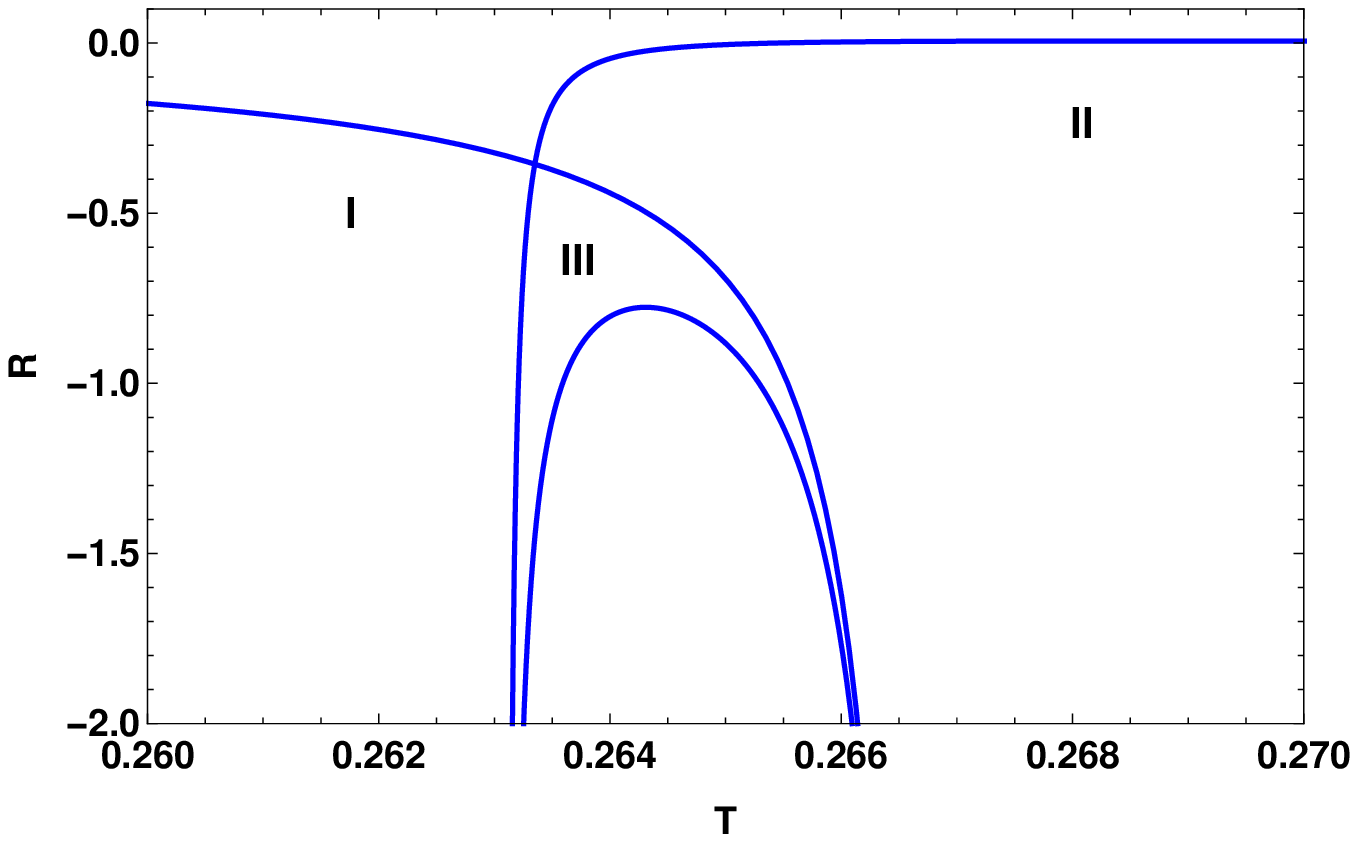}
\end{minipage}%
\begin{minipage}[b]{0.45\linewidth}
\includegraphics[width =2.8in,height=1.8in]{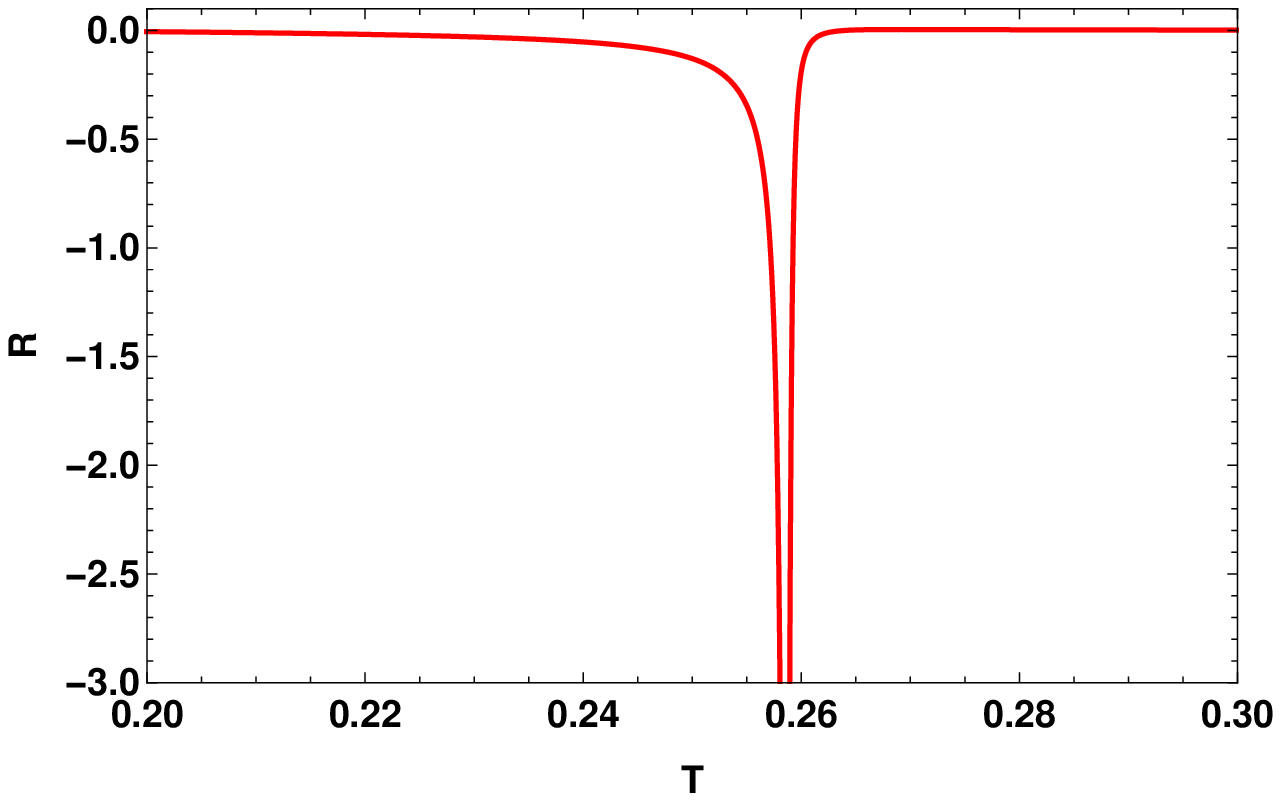}
\end{minipage}\quad
 \caption{\label{Rrnne}\textbf{ Thermodynamic scalar curvatures $R$  of the  RN-AdS black hole in the normal phase space for different Values of $Q$:} Figure (a) is plotted for $Q=0.155$, which exhibits first order phase transition between SBH/LBH . Figure (b) is plotted at  the critical charge $Q=Q_{c}=0.17$, which illustrates the divergence of  $R$ for the  second order critical point.}
\end{figure}
The behaviour of the thermodynamic scalar curvature $R$ with the temperature $T$  are plotted in the fig.[\ref{Rrnne}] parametrically with the radius of the horizon $r_+$ as a parameter. It is observed that the thermodynamic scalar curvature $R$ is a multiple valued function of the thermodynamic variables in the neighbourhood of the first order phase transition similar to the free energy $F$ as shown in fig.[\ref{RN-AdS1}]. As mentioned earlier, the first order phase transition is characterized  through the {\it $R$-crossing Method}, which is depicted in the fig.[\ref{Rrnne}(a)] for the value of the charge $Q < Q_c=0.17$. The crossing of  the two branches $I$ and $II$ indicate  a first order phase transition between SBH/LBH and the transition temperature matches well with that obtained from the free energy consideration but is distinct as the two approaches are quite different. For $Q=Q_c$ the thermodynamic scalar curvature $R$ diverges describing a critical point of second order phase transition as shown in the fig.[\ref{Rrnne}(b)] with the corresponding critical temperature as $T_{c}=0.257$. 
\subsection{Thermodynamic Geometry in the Extended Phase Space}
In the extended phase space the thermodynamic scalar curvature $R$ is obtained from the metric ([\ref{metric rn}) with the free energy $F$ and the temperature $T$ as given by the  Eq.(\ref{RNEFE}). The variables which serve as the coordinates in this case are  $x^{\mu}=(T,\phi)$. The thermodynamic scalar curvature $R$ may then be expressed as follows
    \begin{align}
\label{eqn:eqlabel}
\begin{split}
R=&N/D,\\
N=&36 Q^6 - 3 Q^4 r_+^2 (21 + 256 P \pi r_+^2) +
 Q^2 r_+^4 (19 + 512 P \pi r_+^2 + 256 P^2 \pi^2 r_+^4)\\
 & - 4 r_+^6 (-1 + 2 P \pi r_+^2 + 16 P^2 \pi^2 r_+^4 + 1024 P^3 \pi^3 r_+^6),\\
D=&3 \pi (-Q^2 + r_+^2 + 8 P \pi r_+^4) (-3 Q^2 r_+ + r_+^3 - 8 P \pi r_+^5)^2.
\end{split}
\end{align} 
\begin{figure}[H]
 \centering
\begin{minipage}[b]{0.45\linewidth}
\includegraphics[width =2.8in,height=1.8in]{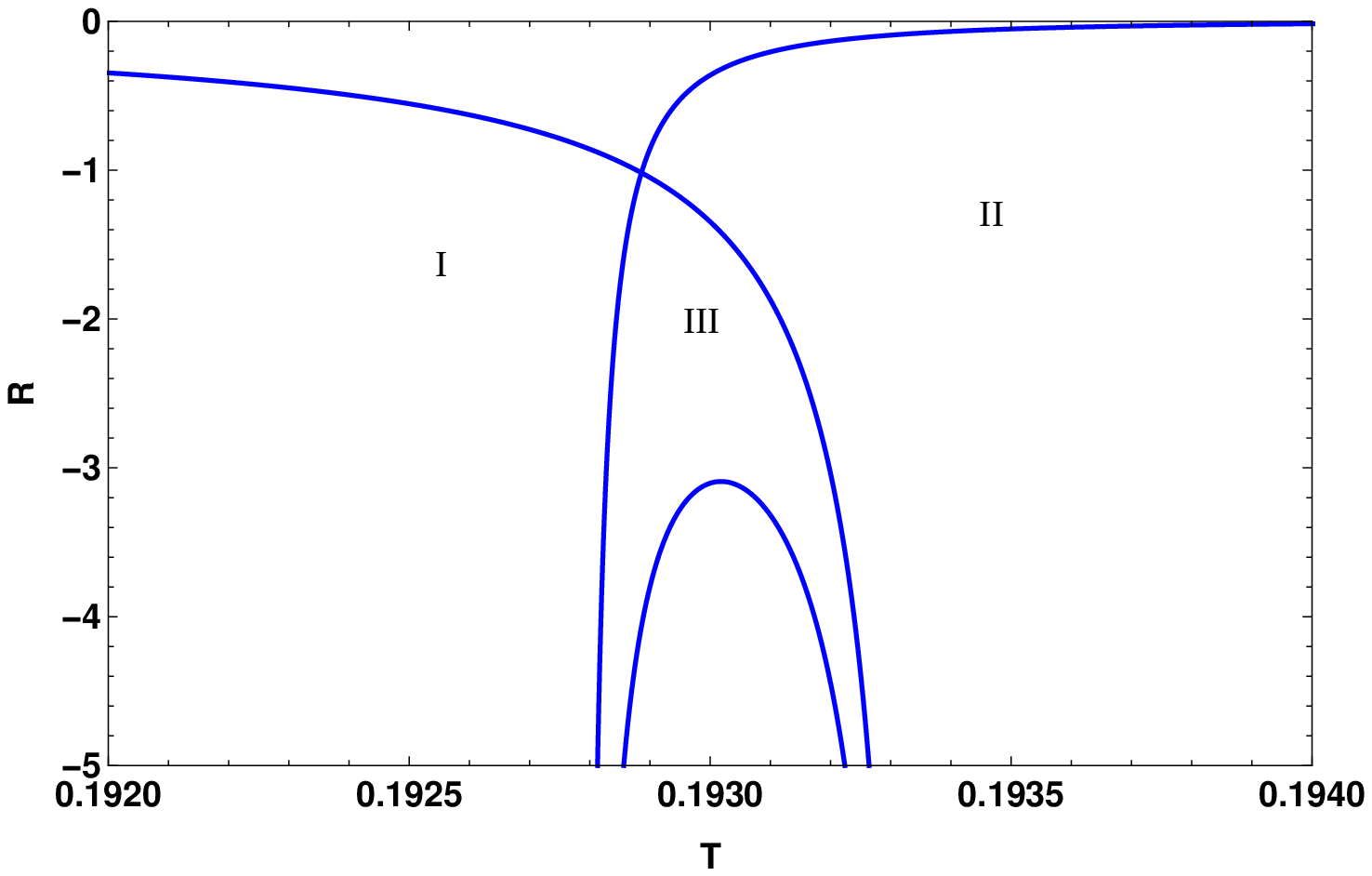}
\end{minipage}%
\begin{minipage}[b]{0.45\linewidth}
\includegraphics[width =2.8in,height=1.8in]{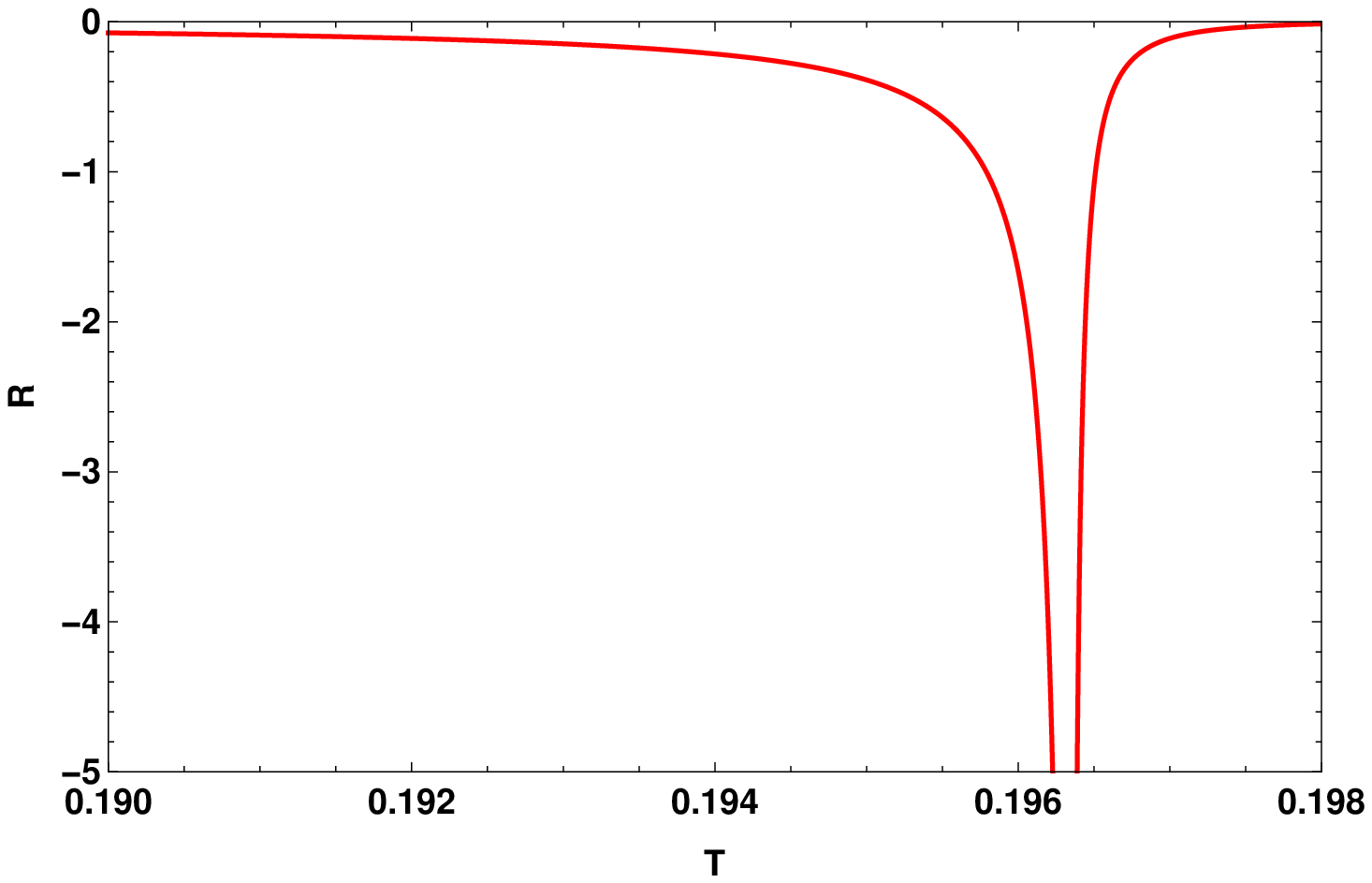}
\end{minipage}\quad
\caption{  \label{rne}\textbf{ Thermodynamic scalar curvature $R$ of the RN-AdS black hole in the extended phase space for fixed values of $P$ and $Q$:} Figure (a) is plotted for $P=0.065$ and $Q=0.22$, which exhibits a first order phase transition between SBH/LBH. Figure (b) is plotted for $P=0.068$ and $Q=0.22$, which illustrates the divergence of $R$ for the second order critical point.}
\end{figure}

As in the previous case the behaviour of the thermodynamic scalar  curvature $R$ with the temperature $T$ for different values of pressure $P$ and a fixed value of the charge $Q=0.22$ are plotted in the fig.[\ref{rne}] parametrically with the radius of the horizon $r_+$ as a parameter.  It is observed from the fig.[\ref{rne}(a)] that for $P<P_{c}$, where $P_{c}$ is the critical pressure the  two different branches $I$ and $II$ of $R$ cross each other indicating a first order phase transition between SBH/LBH. The curve for the  critical pressure $P=P_{c}=0.68$ is illustrated in the fig [\ref{rne}(b)], where the thermodynamic scalar curvature $R$ diverges describing a critical point of second order phase transition with the corresponding critical temperature as $T_{c}=0.196$.
\section{Thermodynamic Geometry of Kerr-AdS Black Holes }
In this section  we construct the thermodynamic geometry of four dimensional Kerr-AdS black holes in the canonical ensemble both in the normal and the extended thermodynamic phase space. As earlier we employ the Helmholtz free energy $F$ to compute the thermodynamic scalar curvature $R$  from the metric Eq.  (\ref{metric rn}). In the canonical ensemble the angular momentum $J$ is held fixed, so the appropriate thermodynamic variables are the temperature $T$ and the angular velocity $\Omega$ for the thermodynamic state space. 

\subsection{Thermodynamic Geometry in the Normal Phase Space}
In the normal phase space the thermodynamic scalar curvature $R$ for four dimensional Kerr-AdS black holes is obtained from the metric defined by the Eq. (\ref{metric rn}), with the free energy $F$ and the temperature $T$ as given by  Eq. (\ref{kerr normal}). The thermodynamic variables in this case are  $x^{\mu}=(T,\Omega)$. The  corresponding thermodynamic scalar  curvature $R$ may be expressed as $R=XN/D$, where 
\begin{align}
\label{eqn:eqlabel}
\begin{split}
N=&A+B+C,\\
X=&6\pi^{\frac{3}{2}} \left({\frac{S}{\pi +S}}\right)^{\frac{1}{2}},\\
A=&3(\pi-2S)S^8(\pi+S)^7+12J^2\pi^3 S^6(\pi+S)^5(28\pi^2+44\pi S+21S^2),\\
B=&16J^4\pi^6 S^4(\pi+S)^3 (66\pi^3+8\pi^2 S-189\pi S^2-126 S^3)\\
&+256J^8\pi^{13}(3\pi^3 -21\pi^2 S-70\pi S^2-48 S^3),\\
C=&64 J^6\pi^9 S^2(\pi +S)(36\pi^4+44\pi^3  -23\pi^2 S^2+12\pi S^3 + 48 S^4),\\
 D=&\left({4J^2\pi^3+S^2(\pi+S)}\right)^{\frac{1}{2}} \\
 &\left[(\pi-3S)S^4(\pi+S)^3-24J^2 \pi^3 S^2(\pi+S)^2 (\pi+2S)-16J^4 \pi^7(3\pi+4S)\right]^2.
\end{split}
\end{align}  
\begin{figure}[H]
   \centering
\begin{minipage}[b]{0.45\linewidth}
\includegraphics[width =2.8in,height=1.8in]{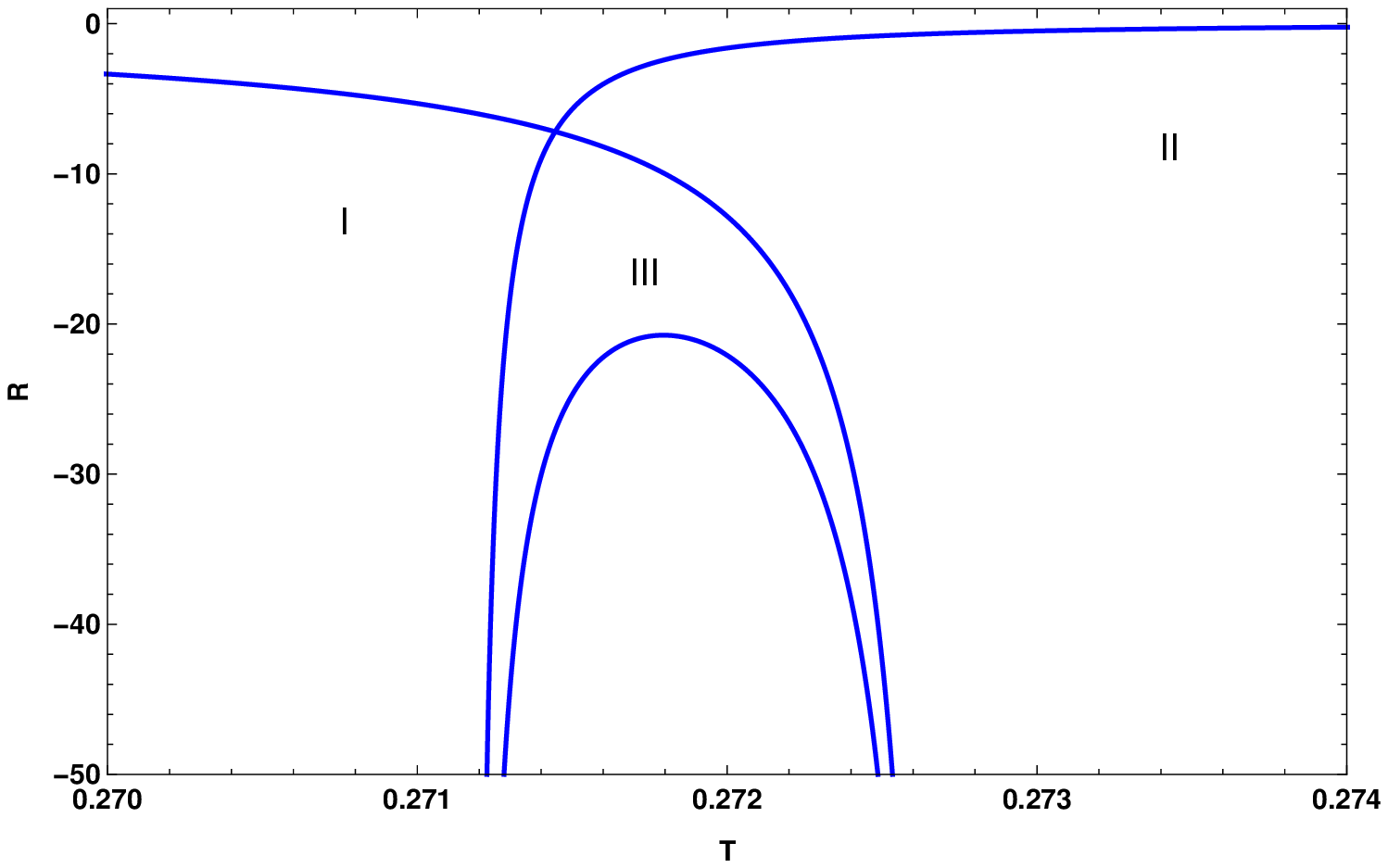}
\end{minipage}%
\begin{minipage}[b]{0.45\linewidth}
\includegraphics[width =2.8in,height=1.8in]{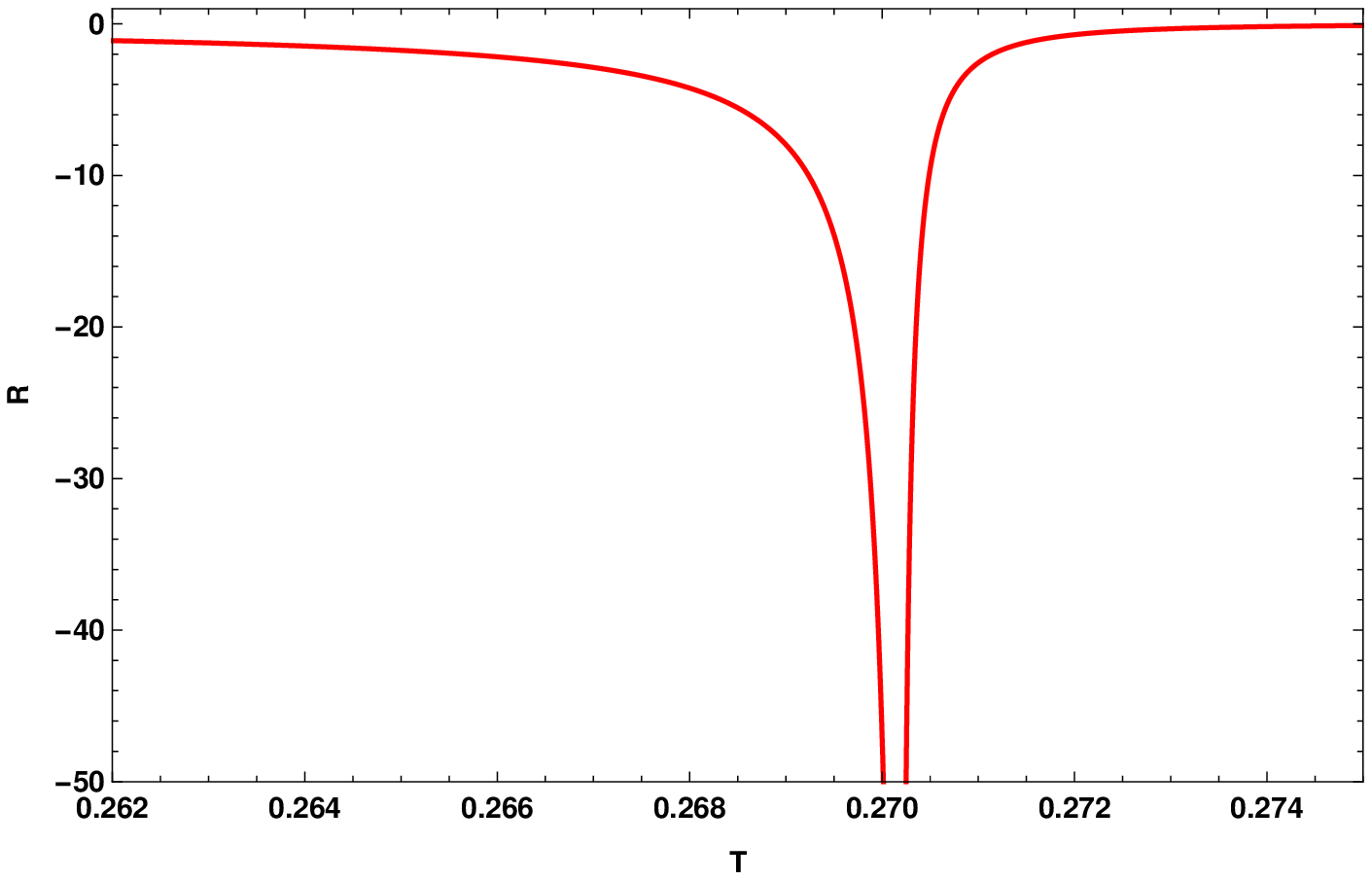}
\end{minipage}\quad
 \caption{\label{kn}\textbf{Thermodynamic scalar curvatures of the Kerr-AdS black hole in the normal phase space  for fixed values of $J$:} Figure (a) is plotted for $J=0.022$, which exhibits a first order phase transition between SBH/LBH. Figure (b) is plotted for the critical angular momentum $J=J_c=0.0236$, which illustrates the  divergence of $R$ for the second order critical point.}
\end{figure}
The behaviour  of  the thermodynamic scalar curvature $R$ with the temperature $T$ for different values of the angular momentum  $J$ are plotted in the fig.[\ref{kn}] parametrically with the entropy $S$ as a parameter. The thermodynamic scalar curvature $R$ in the  fig.[\ref{kn}](a) is plotted against the temperature $T$, for $J=0.022$. This once again exhibits a  first order phase transition between SBH/LBH through the {\it $R$-Crossing Method} where the two branches $I$ and $II$ for $R$ cross each other. The fig.[\ref{kn}](b) is plotted for  $J=J_{c}=0.0236$, where $J_{c}$ is the critical angular momentum where the thermodynamic scalar curvature diverges indicating a second order  phase transition with the corresponding critical temperature as $T=T_c=0.270$ .
\subsection{Thermodynamic Geometry in the Extended Phase Space}
In the extended phase space the  thermodynamic scalar curvature $R$ for four dimensional Kerr-AdS black holes is obtained as earlier from  the metric Eq. (\ref{metric rn}) with the free energy $F$ and the temperature $T$ as given by the Eq. (\ref{kerr extended}). The thermodynamic variable in this case are
$x^{\mu}=(T,\Omega)$ where the pressure $P$ is held fixed. The thermodynamic scalar curvature $R$ may then be expressed as $R=XN/D$  where
\begin{align}
\label{eqn:eqlabel}
\begin{split}
N=&A+B+C,\\
X=&-6\left({S \pi(}{\frac{3}{P}+8S})\right)^{\frac{1}{2}},\\
A=&S^8(3+8PS)^7(-3+16PS)-144J^2\pi^2S^6(3+8PS)^5(21+8PS(11+14PS)),\\
B=&864 J^4 \pi^4 S^4 (3 + 8 P S)^3 \left[-99 +  32 P S (-1 + 7 P S (9 + 16 P S))\right]\\
   & + 62208 J^8 \pi^8 (-27 + 8 P S (63 + 16 P S (35 + 64 P S))),\\
   C=&-6912 J^6 \pi^6 S^2(3+8PS)243+8PS[99+2PS(-69+32PS(3+32PS))],\\
   D=&(3+8PS)\sqrt{\frac{(12J^2\pi^2+S^2(3+8PS)}{P}}(S^4(-1+8PS)(3+8PS)^3\\ 
      &+24J^2\pi^2S^2(3+8PS)^2(3+16PS)+144J^4\pi^4(9+32PS))^2.
\end{split}
\end{align}

\begin{figure}[H]
\centering
\begin{minipage}[b]{0.45\linewidth}
\includegraphics[width =2.8in,height=1.8in]{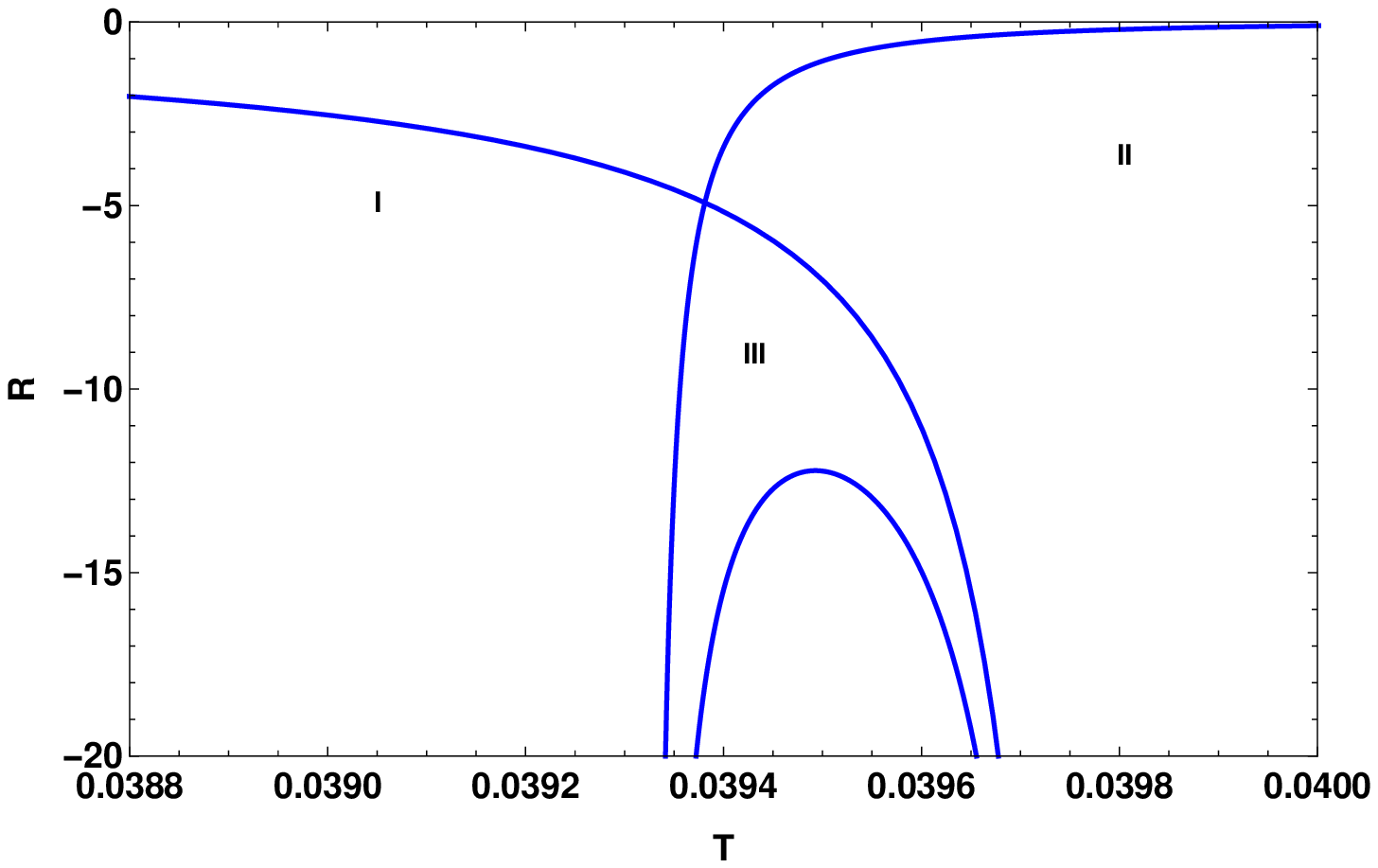}
\end{minipage}%
\begin{minipage}[b]{0.45\linewidth}
\includegraphics[width =2.8in,height=1.8in]{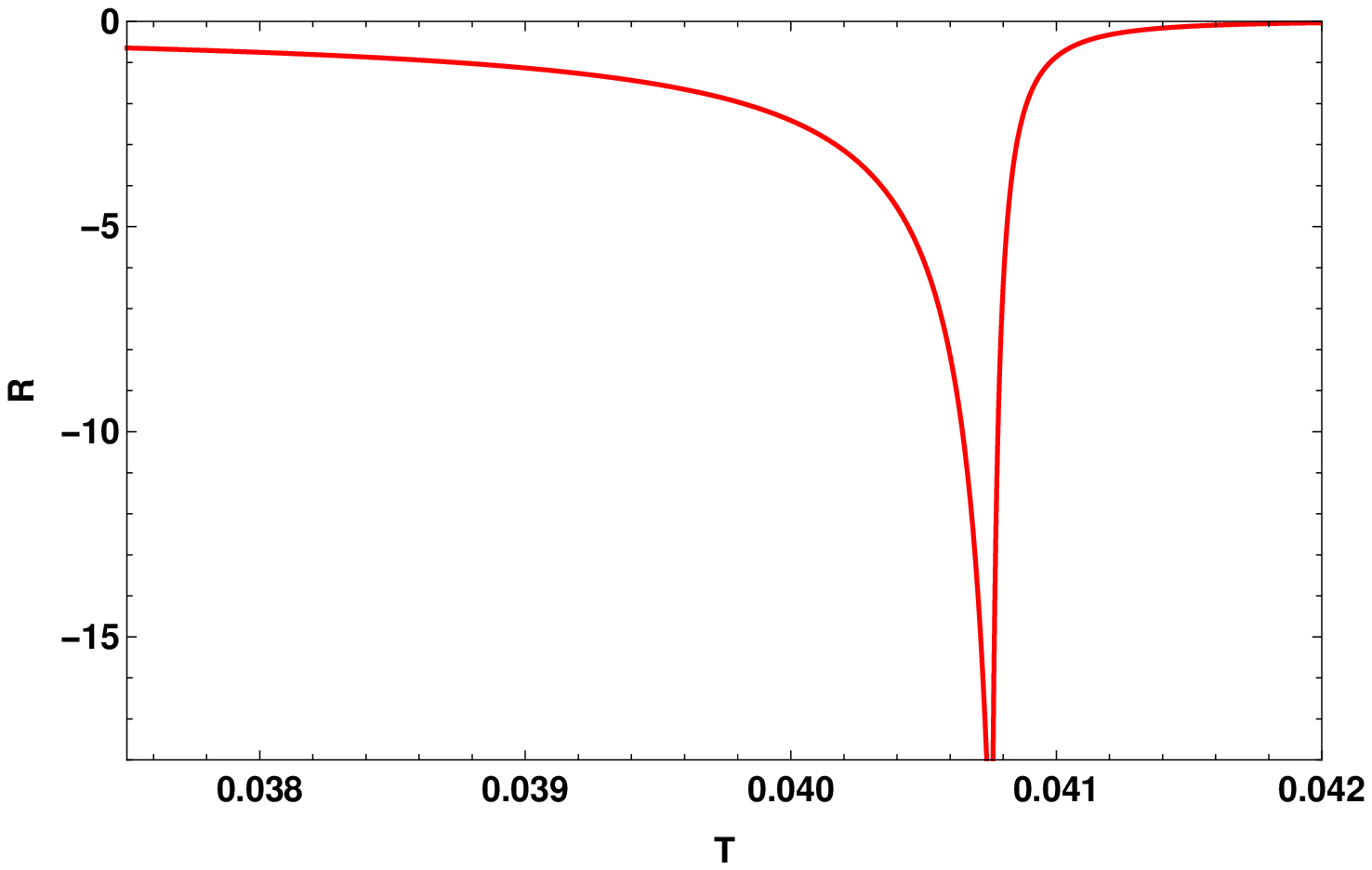}
\end{minipage}\quad
\caption{\label{ke}\textbf{Thermodynamic scalar curvatures of the Kerr-AdS black hole in the extended phase space for a fixed value of  $P$ and $J$:} Figure (a) is plotted for $P=0.0025$ and $J=1$, which exhibits a first order phase transition between SBH/LBH.  Figure (b) is plotted for the critical pressure $P=P_c=0.0027$ and $J=1$, which illustrates the divergence of $R$ for the   second order critical point. }
\end{figure}

As in the previous case the behaviour  of the thermodynamic scalar  curvature $R$ with the temperature $T$ for  different pressures  $P$ and  a fixed angular momentum  $J=1$ are plotted in fig.[\ref{ke}] parametrically with the entropy $S$ as a parameter. As earlier the fig.[\ref{ke}]$(a)$ indicates a first order phase transition between SBH/LBH through the $R$-crossing of the corresponding branches where the pressure $P=0.0025$.
The fig.[\ref{ke}]$(b)$ is plotted for the critical pressure $P=P_{c}=0.0027$, indicates a second order phase transition through the divergence of $R$ at the corresponding critical temperature $T_c=0.040$.
\section{Summary and Discussions}
In summary we have investigated  the thermodynamics and critical phenomena for four dimensional RN-AdS and Kerr-AdS black hole in the canonical ensemble employing the framework of thermodynamic geometry both for the normal and the extended phase (state) space. We emphasize here that the construction of the thermodynamic geometry for black holes in the canonical ensemble was an unresolved issue in this area which precluded the geometrical characterization of the corresponding phase structure
for black holes in this ensemble (see also \cite{sahay2016state}). Through a careful analysis of this outstanding issue we have determined the appropriate thermodynamic potential and the variables for the construction of the thermodynamic metric 
for these black holes, specific to the canonical ensemble both for the normal and the extended thermodynamic phase space.
It has been shown that the thermodynamic scalar curvature $R$ obtained from this metric clearly describes the phase structure for these black holes in the canonical ensemble which conforms to the established unified geometrical characterization for phase transitions described in the literature. 

For the normal phase space where the cosmological constant is held fixed we have shown that the thermodynamic scalar curvature $R$ exhibits the crossing of its branches corresponding to the two coexisting phases at a  first order liquid-gas like phase transition which is analogous to the van der Waals fluid. Interestingly this exactly conforms to the $R$-Crossing Method proposed by one of the authors earlier. The sequence of first order sub critical phase transitions culminate in a critical point describing a second order phase transition at which the thermodynamic scalar curvature $R$ diverges as a function of the temperature. The critical temperature thus obtained matches very well with that obtained from the conventional free energy approach.

In the extended thermodynamic phase space the cosmological constant $\Lambda$ is identified as the thermodynamic pressure
with a corresponding conjugate thermodynamic volume. This naturally leads to the modification of the fist law where the variation of the thermodynamic pressure now needs to be included. As a consequence the ADM mass of the black hole must now be identified with the enthalpy instead of the internal energy as was the case for the normal phase space. We have shown that the thermodynamic scalar curvature  correctly characterizes the phase structure and the critical point 
both for the RN-AdS and the Kerr-AdS black holes in the canonical ensemble for the extended thermodynamic phase space.

We emphasize here that the present article is a further confirmation of the unified geometrical approach to phase transitions and supercritical phenomena in particular the {\it $R$-Crossing Method} as applied to black holes considered as thermodynamic systems. It is important to further elucidate the application of this geometrical framework to study the phase structures and critical phenomena for other black holes like the Kerr-Newman-AdS black holes in the canonical ensemble both for the normal and the extended phase space. In this context it would be an interesting issue to explore the supercritical regime and understand the significance of the Widom line for black holes. We leave these interesting issues for future explorations.
\section{Acknowledgement}
We would like to thank Anurag Sahay for many useful and stimulating discussions.
\bibliographystyle{unsrt}
 \bibliographystyle{ieeetr}
\bibliography{EE_RNbib}
\end{document}